\documentclass[12pt]{iopart}
\usepackage{graphicx}

\usepackage{iopams}
\usepackage{subfig}
\begin{document}

\title[Suppression of entanglement in two-mode Gaussian open systems
]{Suppression of entanglement in two-mode Gaussian open systems}

\author{Tatiana Mihaescu and Aurelian Isar}

\address{Department of Theoretical Physics, National Institute of Physics and Nuclear Engineering,
P.O.Box MG-6, Bucharest-Magurele, Romania}
\ead{mihaescu.tatiana@theory.nipne.ro, isar@theory.nipne.ro}

\begin{abstract}
We study the evolution of the entanglement of two independent bosonic modes embedded in a thermal environment, in the framework of the theory of open quantum systems. As a measure of entanglement we use the logarithmic negativity. For a non-zero temperature of the thermal reservoir the entangled initial Gaussian states become always separable in a finite time. For initial squeezed thermal states we calculate the survival time of entanglement and analyze its dependence on temperature, squeezing parameter and mean thermal photon numbers. For a zero temperature of the thermal bath an entangled initial state remains entangled for all finite times, but in the limit of asymptotically large times it becomes separable.
\end{abstract}


\section{Introduction}

The theory of quantum information has obtained recognition through studying quantum correlations like quantum entanglement, which enabled interesting applications in quantum information processing. In particular, the study of quantum entanglement in continuous variable systems using Gaussian states is one of the most relevant examples \cite{bra1}. Two-mode Gaussian systems are completely described by the covariance matrix, from which symplectic invariants can be obtained. In the case of open systems, the irreversible dissipative evolution is determined by completely positive quantum dynamical semigroups. The time evolution of a system of two independent initially entangled bosonic modes embedded in a thermal environment, described by the master equation, leads to the suppression of the quantum entanglement. We calculate the survival time of entanglement for initial squeezed thermal states and the obtained expressions show a strong dependence of the survival time on the squeezing  parameter, temperature of the thermal bath and the mean photon numbers. For a non-zero temperature of the reservoir an entangled initial squeezed thermal state always becomes separable, but in the case of a zero temperature of the environment the state remains entangled for all finite times.

\section{Evolution of a two-mode system interacting with a thermal bath}

The irreversible dissipative evolution of an open quantum system can be described by a completely positive quantum dynamical semigroup, under the Markovian approximation that the system has a weak local interaction with the environment. Namely, the evolution of the open system is described by the following general Kossakowski-Lindblad master equation for the density operator $\rho$ \cite{gor,lin,rev}:
\begin{equation}
\frac {d\rho(t)} {dt}=-\frac i \hbar [H, \rho(t)]+\frac {1} {2\hbar}\sum_k \big(2 B_k\rho(t) B_k^\dag-\{\rho(t),B_k^\dag B_k\}_+\big),
\label{masteq}\end{equation}
where $H$ is the Hamiltonian of the system and $B_k$ are operators which determine the interaction of the system with the environment. In the case of a two-mode bosonic system with the Hamiltonian
\begin{equation}
H=\frac{1}{2m}(p_x^2+p_y^2)+\frac{m}{2}(\omega_1^2 q_x^2+\omega_2^2 q_y^2),
\end{equation}
where $m$ is the mass and $\omega_1$, $\omega_2$ are the single mode frequencies,
the Gaussian form of the states is preserved during the interaction with the environment if one takes the operators $B_j$ as polynomials of the first degree in the canonical variables of coordonates $q$ and momenta $p.$
The covariance matrix of the bimodal system is given by
\begin{eqnarray}
\sigma(t)=
\left(\matrix{\sigma_{q_x q_x}(t)&\sigma_{q_x p_x}(t) &\sigma_{q_x q_y}(t)&
\sigma_{q_x p_y}(t)\cr \sigma_{q_x p_x}(t)&\sigma_{p_xp_x}(t)&\sigma_{q_yp_x}(t)
&\sigma_{p_xp_y}(t)\cr \sigma_{q_xq_y(t)}&\sigma_{q_yp_x}(t)&\sigma_{q_yq_y}(t)
&\sigma_{q_yp_y}(t)\cr \sigma_{q_xp_y}(t)&\sigma_{p_xp_y}(t)&\sigma_{q_yp_y}(t)
&\sigma_{p_yp_y}(t)}
\right),
\end{eqnarray}
where the elements defined by
\begin{equation}
\sigma_{AB}(t)= \frac 1 2 \Tr[\rho(AB+BA)(t)]-\Tr[\rho A(t) ] \Tr[\rho B(t) ],
\end{equation}
characterize the correlation between operators $A$ and $B$. From the master equation (\ref{masteq}) one obtains a differential equation describing the evolution of the covariance matrix, which has the following time-dependent solution \cite{san}:
\begin{equation}
\sigma(t)=M(t)(\sigma(0)-\sigma(\infty)) M^{\rm T}(t)+\sigma(\infty),
\end{equation}
with $M(t)=\exp{(Y t)},$ where \cite{san}
\begin{equation}
Y=\left(\matrix{-\lambda &\frac 1 m &0&0\cr
-m\omega_1 &-\lambda&0&0
\cr 0&0&-\lambda&\frac 1 m
\cr 0&0&-m\omega_2
&-\lambda}\right)
\end{equation}
and $\lambda$ is the dissipation constant. $\sigma(\infty)$ is the
covariance matrix corresponding to the asymptotic Gibbs state of the
two bosonic modes in thermal equilibrium at temperature $T.$ Its
diagonal elements are given by \cite{rev}:
\begin{eqnarray}m\omega_1
  \sigma(\infty)_{q_xq_x}=\frac{\sigma(\infty)_{p_xp_x}}
  {m\omega_1}=\frac 1 2 \coth(\frac{\omega_1}{2 k T}),\\
m\omega_2 \sigma(\infty)_{q_yq_y}=\frac{\sigma(\infty)_{p_yp_y}}{m\omega_2}=\frac 1 2 \coth(\frac{\omega_2}{2 k T}),\end{eqnarray} where $k$ is the Boltzmann constant, while all the non-diagonal elements are 0.

\section{Dynamics of continuous variable entanglement}

For two-mode Gaussian states the covariance matrix is a $4 \times 4$ real, positive and symmetric matrix, written in block form as follows:
\begin{equation}
\sigma(t)=\left(\matrix{
A & C \cr
C^{\rm {T}} & B}
\right),
\end{equation}
where $A$ and $B$ are the covariance matrices for the single modes, and $C$ denotes the cross-correlations between the modes.
By means of local unitary operations, the first statistical moments can be set to zero, without affecting the entanglement.
A well known condition of separability of a state is the PPT (positive partial transpose) criterion, which is proved to be also sufficient for Gaussian states \cite{sim}.
Using this criterion, one can introduce the logarithmic negativity as a measure of entanglement (inseparability), which can be written in terms of the symplectic invariants \cite{ser4,aijqi,aosid}:
\begin{eqnarray}
E_{N}(t)=-\log_2[(\det A +\det B)\nonumber\\-2\det C
-2\sqrt{\left[\frac{1}{2}(\det A+\det B)-\det
C\right]^2-\det\sigma(t)}].
\end{eqnarray}

\begin{figure}
 \subfloat[~Logarithmic negativity $E_N$ versus time $t$ and squeezing parameter $r$ for an entangled initial squeezed vacuum state and $c\equiv\coth{\frac {\omega}{2kT}}=2$, $\omega_1=\omega_2\equiv \omega, m=\hbar=\lambda=1$.
]{{\includegraphics[width=0.5\textwidth]{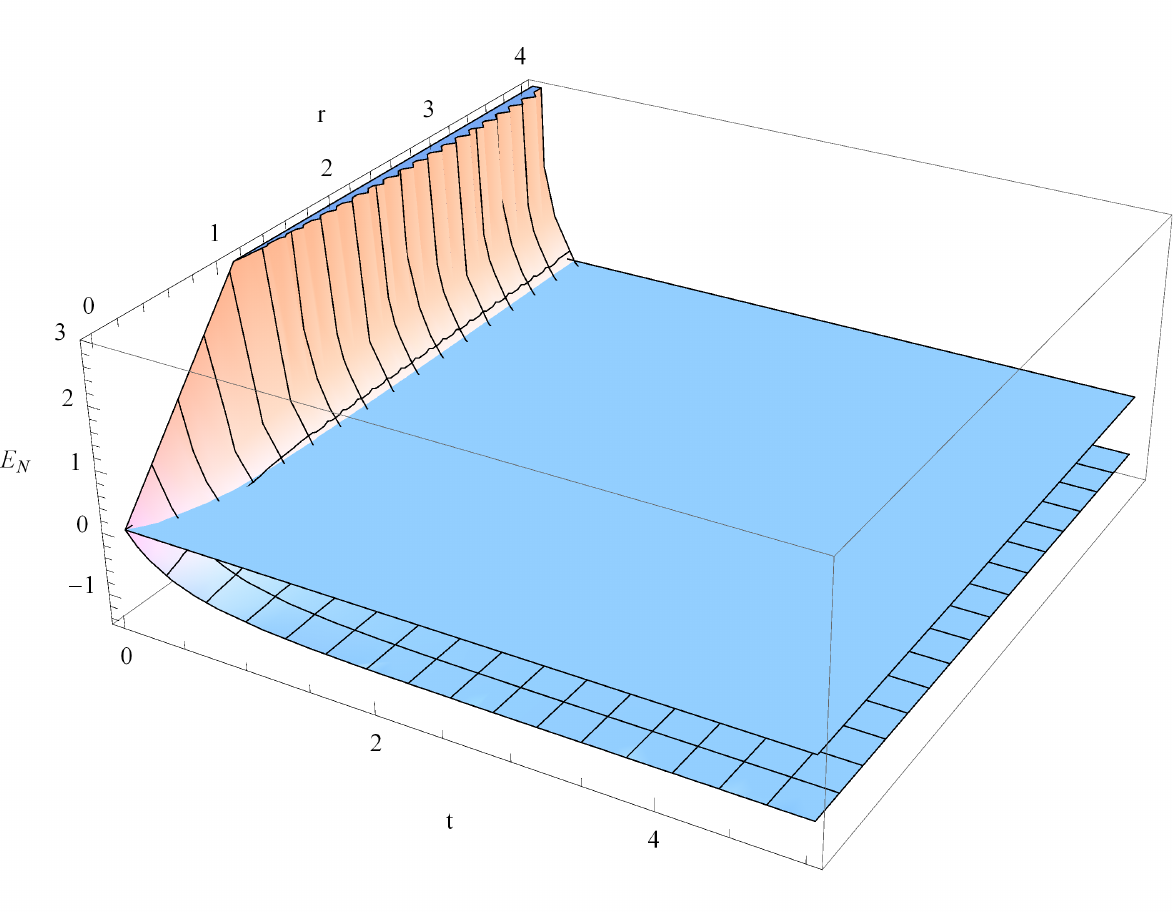} }}
\label{fig:1}
\qquad
 \subfloat[~Logarithmic negativity $E_N$ versus time $t$ and temperature (through $c\equiv\coth{\frac {\omega}{2k T}}$) for an entangled initial squeezed vacuum state and $r=2$,  $\omega_1=\omega_2\equiv \omega, m=\hbar=\lambda=1$.
\label{fig:2}]{{\includegraphics[width=0.5\textwidth]
{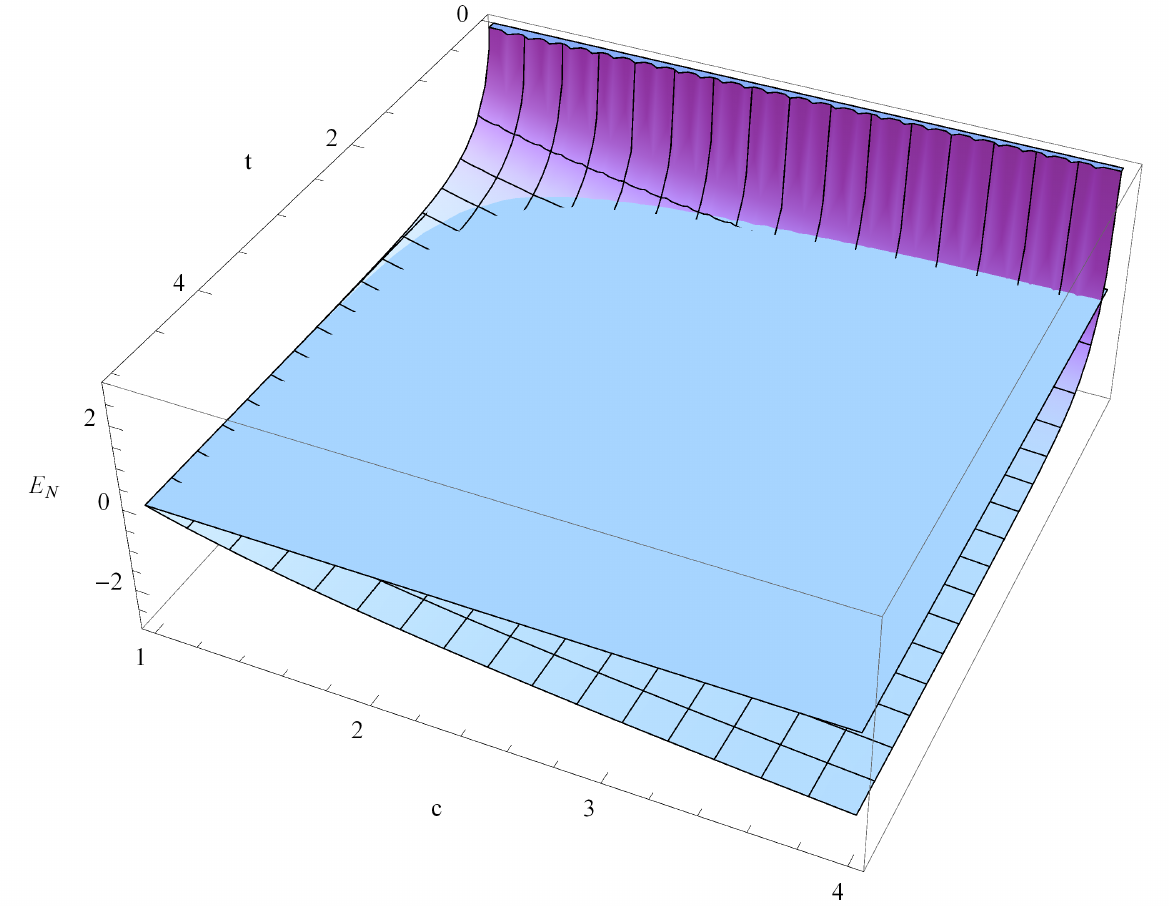}}}
\caption{Dependence of logarithmic negativity.}
\end{figure}

In order to describe the dynamics of the entanglement we consider an entangled initial squeezed thermal state, with the covariance matrix of the form:
\begin{equation}
\sigma(0)=\left(\matrix{
a & 0 & c & 0 \cr
0 & a & 0 & -c \cr
c & 0 & b & 0 \cr
0 & -c & 0 & b}
\right),
\end{equation},
\begin{eqnarray}
a&=&n_1 \cosh^2 r +n_2 \sinh^2 r + \frac 1 2 \cosh 2r, \\
b&=&n_1 \sinh^2 r +n_2 \cosh^2 r + \frac 1 2 \cosh 2r, \\
c&=&\frac 1 2 (n_1+n_2+1)\sinh 2r,
\end{eqnarray}
where $n_1$, $n_2$ are the mean photon numbers of the two modes and $r$ is the squeezing parameter. The state is entangled if $r>r_s,$, where $\cosh^2r_s= (n_1+1)(n_2+1)/(n_1+n_2+1)$ \cite{mar}.

The entanglement evolution of the squeezed vacuum state ($n_1=n_2=0$) at non-zero equilibrium temperature $T$ is illustrated in Fig. 1(a), where it is represented the logarithmic negativity as a function of time $t$ and squeezing parameter $r$. An entangled initial state becomes separable at a certain moment of time, this phenomenon being known as entanglement sudden death. In Fig. 1(b) we show the logarithmic negativity dependence on time $t$ and temperature, for an initial squeezed vacuum state. One can observe that for zero temperature ($c=1$) the entangled initial squeezed vacuum state becomes separable in the limit of large times, corresponding to an asymptotic Gibbs product state.

\subsection{Survival time of entanglement}

By imposing the logarithmic negativity to be zero, we obtain the following expression for the survival time of entanglement, in the particular case of a symmetric initial state with $n_1=n_2=n$, $\omega_1=\omega_2=1$ and $m=\lambda=1$, as a function of the squeezing parameter $r$ and temperature $T$ ($c\equiv\coth{\frac {\omega}{2k T}}$):
				\begin{equation}
t_s=\frac 1 2 \ln[\frac{c\ e^{2r}-1-2n} { e^{2r}(c-1)}].\label{temp}
										\end{equation}
From Eq. (\ref{temp}) we can deduce that at the temperature $T=0$ ($c=1$), the state becomes separable in the limit of asymptotically large time. This is also seen from Fig. 2 (a), where it is represented the dependence of the survival time of entanglement on squeezing parameter and temperature. For $r<r_s$ the state is separable, and for a temperature approaching $T=0$, the survival time of entanglement is rapidly increasing. In Fig. 2 (b) it is shown the dependence of the survival time on the mean photon number $n$ and squeezing parameter $r$. As expected, one can see that for small mean photon numbers and for a given squeezing parameter, the state remains entangled for a longer time.  Also, from all the displayed graphics we observe that, by increasing  $r,$$T$ and $n$, the survival time of entanglement tends to a constant value.
	
In another particular case when we fix the temperature
                ($c=2$) and the mean photon numbers ($n_1=n_2=1$), we
                obtain the following dependence of the survival time
                on the squeezing parameter $r$ and frequency
                $\omega_1=\omega_2=\omega$:
\newpage

\begin{eqnarray}
t_s=\frac 1 2 \ln[\frac {1}{3 \omega}
(4\omega-3e^{-2r}(\omega^2+1)\nonumber
\\+e^{-2r}\sqrt{9(\omega^4-\omega^2+1)-2e^{2r}\omega(3\omega^2-2 e^{2r}\omega+3)})].
\end{eqnarray}
The calculations show that the survival time depends very slightly on the frequency.

\begin{figure}
 \subfloat[~Survival time of entanglement $t_s$ versus squeezing parameter $r$ and $c=\coth{\frac {\omega}{2k T}}$ for $n_1=n_2=1$ and $\lambda=m=1$.
]{{\includegraphics[width=0.5\textwidth]{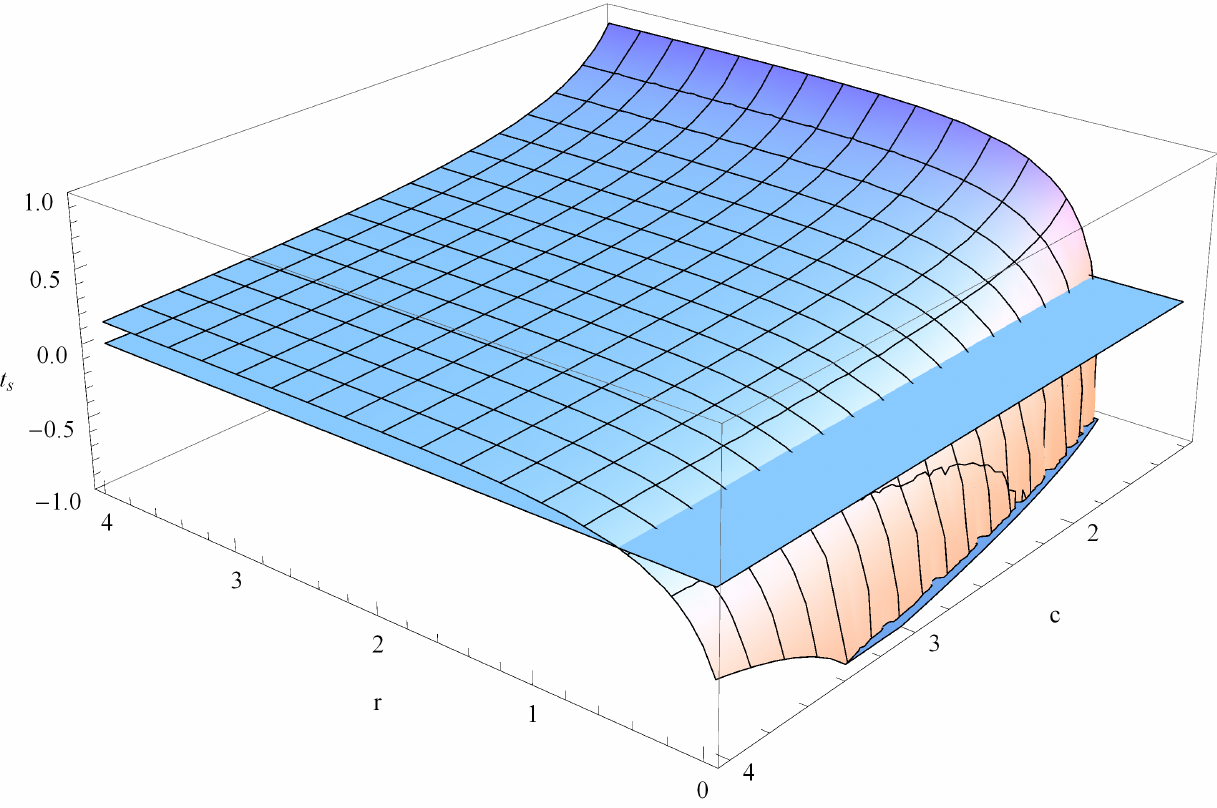} }}
\label{fig:3}
\qquad
 \subfloat[~Survival time of entanglement $t_s$ versus squeezing parameter $r$ and $n_1=n_2=n$ for $c=\coth{\frac {\omega}{2k T}}=2$ and $\lambda=m=1$.
\label{fig:4}]{{\includegraphics[width=0.5\textwidth]
{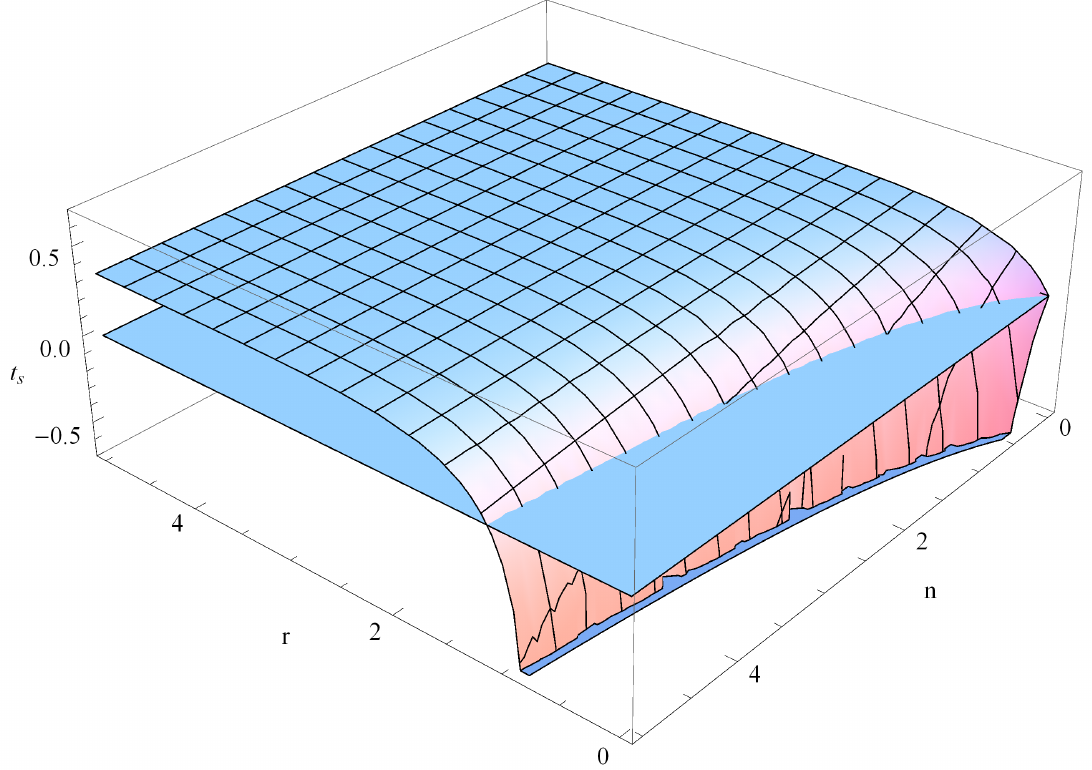}}}
\caption{Dependence of the survival time of entanglement.}
\end{figure}

The obtained expressions for the survival time of entanglement are in agreement with the previously obtained results \cite{ros}. For a zero temperature of the reservoir the two bosonic modes remain entangled for all finite times, but their state becomes separable for asymptotically large times. For a non-zero temperature of the bath, the survival time of entanglement depends on the initial state of the system (squeezing parameter and mean photon numbers), dissipation constant and temperature.

\end{document}